\documentstyle[12pt]{article}
\begin{document}

\renewcommand{\thefootnote}{\fnsymbol{footnote}} \thispagestyle{empty}

\begin{flushright}
JINR E2-2000-207
\end{flushright}

\begin{flushright}
IC/2000/149
\end{flushright}

\vspace{.5cm}

\begin{center}
{\large {\bf Anomalous $U(1)_{A}$ and Electroweak Symmetry Breaking}} 
\vspace{1cm}\\[0pt]

Ilia~ Gogoladze${}^{a,c}$
\footnote{ e-mail: iliag@ictp.trieste.it}, 
~and ~~Mirian~ Tsulaia${}^{a,b,c}$ 
\footnote{ e-mail: tsulaia@thsun1.jinr.ru}\vspace{.1cm}\\[0pt]

\vspace{1cm} ${}^a${\it International Centre for Theoretical Physics}\\[0pt]
{\it Trieste, 34100, Italy}\\[0pt]
\vspace{0.5cm} ${}^b${\it Bogoliubov Laboratory of Theoretical Physics, JINR}
\\[0pt]
{\it Dubna, 141980, Russia}\\[0pt]
\vspace{0.5cm} ${}^c${\it The Andronikashvili Institute of Physics, Georgian
Academy of Sciences,}\\[0pt]
{\it Tbilisi, 380077, Georgia} \vspace{1.5cm}\\[0pt]
{\bf Abstract}
\end{center}

We suggest a new mechanism for electroweak symmetry breaking in the
supersymmetric Standard Model. Our suggestion is based on the presence of an
anomalous $U(1)_{A}$ gauge symmetry, which naturally arises in the four
dimensional superstring theory, and heavily relies on the value of the
corresponding Fayet--Illiopoulos $\xi$--term.

\vfill 
\setcounter{page}0
\renewcommand{\thefootnote}{\arabic{footnote}} \setcounter{footnote}0
\newpage

\section{Introduction}

Recently the theories suggesting the fundamental
scale of gravitational
interaction to be as low as a few TeVs
were developed \cite{add}. Within
these theories the observed weakness of the
Newtonian coupling is due to the
existence of large \ ($\gg $TeV$^{-1}$) \ extra
dimensions into which the
gravitation flux can spread out. At the distances
larger than the typical
size of these extra dimensions gravitational
potential goes to its standard
Newton's law. Moreover, all currently known
collider data as well as
astrophysical and cosmological observations are
barely consistent with the
known theoretical construction in the case of
two extra dimensions, at least
\cite{cons}.

A number of recent publications consider the
origin of electroweak symmetry
breaking within the low--scale extra
dimension scenario. 
Several origins of electroweak
symmetry breaking and  of the Higgs
field were suggested. For instance, in \cite{comp} Higgs field
was considered as a
composite particle while in \cite{rad7} Higgs field
was identified with a
tree--level massless open string state.
In \cite{strong5} the
spontaneous electroweak symmetry breaking by
supersymmetric strong dynamics
at the TeV scale was studied. Also in \cite{dpq3} the
issue of extra
dimensions at the TeV scale as a possible origin
of electroweak symmetry
breaking was suggested.

In the present paper we address the issue of
the possibility of electroweak
symmetry breaking triggered by the
$D_{A}$--term of anomalous \ $U(1)_{A}$ \
symmetry along with the standard F--terms
coming from the $N=1$
superpotential. To motivate this kind of
scenario we can refer to the
superstring theory. Until the recent time
the compactification of
heterotic \ $E_{8}\otimes E_{8}$ \ superstring on
Calabi--Yau manifolds \cite{CHSW} or alternatively on
the orbifolds \cite{nar} was considered as the
only way to produce the realistic chiral \ $N=1$
\ supergravity theory in four dimensions.
However, the recent developments
in the string theory has revealed that 
other superstring theories like
the type {\bf IIB} and {\bf I/I$^{\prime }$} are phenomenologically
acceptable as well \cite{K} -- \cite{AKT}.
 In latter theories as well as in the case of heterotic string
\cite{BO}, the
string scale may not be directly
related to the Planck scale \cite{type1},
and, hence, the gauge hierarchy problem is naturally avoided. The
compactification of the type {\bf IIB}
(or type {\bf I/I$^{\prime }$})
theory using various kind of
orientifold constructions leads to the string
vacua which contain D--branes.
The gauge fields live on the
D--branes while gravity still propagates in the bulk.
Such kinds of string
vacua contains the spectrum of the
Supersymmetric Standard Model along with
some modulus fields, dilaton, axion,
and the fields from the "hidden sector"
(see \cite{ibq} for a review). The $SU(2)_{W}\otimes U(1)_{Y}$
gauge symmetry of the Standard Model is
enlarged in these kinds of theories
by one or several anomalous $U(1)_{A}$
gauge symmetries. It was shown that
the corresponding Fayet--Illiopoulos $D_{A}$--term
can play  crucial role
in spontaneous supersymmetry breaking \cite{PD},
as well as in a breaking of
some nonanomalous gauge symmetries \cite{FINQ}
and in the strong CP--violation
problem \cite{NILLES}. Therefore it is
interesting to investigate wether the
anomalous $D_{A}$--term has some other
 impact on the four--dimensional
physics namely on electroweak symmetry breaking.
We show that at least for
the case of one anomalous $D_{A}$--term
such a breaking can take place.

The paper is organized as
follows: In Section 2 we recall some basic facts
about the origin of anomalous $U(1)_A$
symmetry and corresponding
Fayet--Illiopoulos \ $\xi$--terms in
four dimensional superstring theories
and we suggest the model where the
electroweak symmetry is broken via the
$D_{A}$--term of anomalous \ $U(1)_{A }$ \
symmetry, without presence of mass
(mixing) terms for Higgs superfields
in four dimensional ($d=4$) \ $N=1$ \
lagrangian. In Section 3 we discuss the
possible dynamical generation of the
$\mu$--term in $d=4$ \, $N=1$ \ SUGRA due
to the condensate of fields from
the "hidden sector" having nontrivial
charge under \ $U(1)_A$ \ symmetry and
examine the consequences of the
appearance of \ $\mu$--term in four
dimensional lagrangian for
the \ $SU(2)_{W} \otimes U(1)_{Y}$ \ symmetry
breaking. Section 4 contains our conclusions and outlook.

\section{The model 1}

\subsection{The origin of anomalous $U(1)_{A}$ symmetry}

The four dimensional field
theoretical models obtained after the
compactification of either heterotic
or type {\bf IIB} superstring theories
contain the anomalous \ $U(1)_{A}$ \ gauge
symmetries. For the case of
compactification of the heterotic
string there exists one anomalous gauge
symmetry \cite{DSW}. The anomalies,
being disastrous for the quantum
consistency of the theory, are cancel
due the four dimensional
Green--Schvartz mechanism, analogous to
the one taking place in ten
dimensional type {\bf I} and heterotic
superstring theories \cite{GS},
therefore, the four dimensional
string theories are consistent as well.
Namely, upon the compactification  on a
six dimensional volume $v$
 the \ $U(1)_{A}$ \ anomalies
for the heterotic string case are
cancel due to the presence in the action
of the counter term  of the form:
\begin{equation}
\delta _{GS}B\wedge F_{{U(1)}},  \label{B2}
\end{equation}
along with the coupling term $dB \wedge \omega$,
where $\delta _{GS}$ is the Green--Schwartz
anomaly cancellation
coefficient, being the universal
constant of the model; \ $F_{{U(1)}}$ \ is
the curvature corresponding to the
anomalous gauge field \ $A_{\mu }$;
$\omega$ is the corresponding Chern--Simons form
\ and \
$B$\ is the four dimensional
antisymmetric two--form which gives the
pseudoscalar \ $a$--axion upon the
dualisation: \ $\partial _{\mu }B_{\nu
\rho }\equiv \epsilon _{\mu \nu \rho \sigma }\partial ^{\sigma }a$\ .
The
combination \ $(s+ia)$ \ of the axion \ $a$ \ and
the heterotic string
dilaton \ $s$ \ forms the lowest component
of the chiral superfield \ $S$
\thinspace , referred to as a complex
dilaton superfield. The latter undergoes
the transformation \ $S\rightarrow S+
\frac{i}{2}\delta _{GS}\theta _{A}$ \
with the parameter \ $\theta _{A}$ \ under the anomalous \ $U(1)_{A}$ \
gauge group. After developing the gauge
invariant kinetic term of the
complex dilaton
\begin{equation}
L_{S}=-\int d^{4}\theta \ln (S+\bar{S}-\delta _{GS}V),  \label{di}
\end{equation}
where \, $V$ \, is the corresponding
vector superfield, one can
see that supersymmetry generates along with (\ref{B2})
the mass term for
anomalous \ $A_{\mu }$ \ gauge boson at
two--loop level and
Fayet--Illiopoulos \thinspace $\xi $--term at
one loop level in string
expansion. The value of
parameter \ $\xi $ \ is given by:
\begin{equation}
\xi =\frac{\delta _{GS}}{4}M_{str}^{2}~~~~~~~\mbox{with}
~~~~~~~\delta _{GS}=
\frac{TrQ_{A}}{48\pi ^{2}},  \label{di2}
\end{equation}
and \thinspace $\sqrt{\xi }$ \thinspace is
of order or less of string scale
depending on the value of \ $~TrQ_{A}$, \ where
the summation in the trace
goes over all charges of superfields
which transform nontrivialy under the
anomalous \ $U(1)_{A}$ symmetry.
The role of Goldstone boson is played by the
axion $a$, which becomes the
longitudinal component of anomalous gauge boson
giving the mass
proportional to \ $\lambda ^{4}\delta _{GS}^{2}M_{str}^{2}$
\ (where \ $\lambda ^{2}=1/\langle s\rangle $ \ is the
heterotic string
coupling constant) and, hence,
anomalous \, $U(1)_{A}$ \,
gauge symmetry is broken spontaneously.

The situation for the case of
compactification to \ $d=4$ \ of the type {\bf IIB}
 superstring theory is
essentially different since there can be several
anomalous \ $U(1)_{A}^{i}$ \ groups.
The quantities \ $\xi ^{i}$ \ are
proportional to the combination of
real--parts of the twisted moduli fields
coming from NS--NS sector with the
model dependent coefficients and vanish
in the orbifold limit. However,
the anomalous \ $U(1)_{A}^{i}$ \ gauge
bosons {\it obtain} the masses of the
order of the string scale despite of whether
the orbifold singularities
are blown up or not \cite{Po}. On the other
hand \ $\xi ^{i}$ \ can be generated
non--perturbatively at the tree level
in string expansion. The order of
parameters \ $\xi ^{i}$ \ is undetermined,
thus, giving the large room for the model building.
In the models considered below we suppose the
 the order of parameters $\xi$ to be close by the magnitude
to the order of type {\bf IIB} string scale \footnote{As we have mentioned
in the Introduction the heterotic string scale can be also lowered
to the 1 TeV due to the effect of small instantons.}
i.e.,  $~\sqrt{\xi }\sim O(10^{2})$ GeV in the low scale
(TeV) string theory.
\ Again,
the four dimensional
type {\bf IIB} superstryng theory is anomaly free,
but now the cancellations
of \quad $U(1)_{A}^{i}$ \quad happens due to
the generalized Green-Schwartz
cancellation mechanism \cite{Po} -- \cite{S} governed
by the imagine parts
of twisted moduli fields coming from the
Ramond--Ramond sector rather than
dilaton as it was for the four dimensional
heterotic string models.

\subsection{The model}

The important problems still open in
the framework of the MSSM are: what is
the origin of \ $\mu$--term, why (how) a
given coupling (or mass) is tiny or
zero without any apparent symmetry reasons,
why the magnitude of \ $\mu$
--term is of the order of \ 100 GeV \ and not
of the order of \ $M_{Str}$ \
which is the fundamental scale in superstring theory.

Therefore we would like to
address the question: {\it Is it possible to
break SM gauge symmetry without $\mu$--term and
without extension of particle spectrum?}

In subsequence we shall not
specify the string origin of the models
considered below. We simply assume that
the lagrangian of the supersymmetric Standard Model 
is enlarged by single anomalous Fayet -- Illiopoulos
$\xi$ - term corresponding to the
 anomalous \ $U(1)_{A}$ \ gauge
symmetry. The $D_{A}$
--term for anomalous \ $U(1)_{A}$ \ has the generic form:
\begin{equation}
D_{A}=\xi +\sum_{i}Q_{i}X_{i}G_{i},  \label{fi}
\end{equation}
where \, $G_{i}$ \, is the derivative of
corresponding K\"{a}hler potential of \ $d=4$
\ $N=1$ \ supergravity with
respect to fields \ $X_{i}$ \ having the charges
\ $Q_{i}$ \ under the anomalous gauge group.
Retaining the quadratic part of
K\"{a}hler potential with respect to
the matter fields one obtains:
\begin{equation}
D_{A}=\xi +\sum_{i}Q_{i}|X_{i}|^{2}.  \label{FI}
\end{equation}
One can see that it is always
possible to define the anomalous $U(1)_{A}$
\, charges of quarks and leptons in
the way that all terms of the ordinary
MSSM lagrangian are allowed
except the \ $\mu $--term. On the other hand we
can also use \ $U(1)_{A}$ \ symmetry
for the suppression of \ R--parity breaking
terms. For example, one of the
possible set of anomalous \ $Q_{i}$\ charges
for one family of quarks and leptons
is given in Table 1. Here \ $Q_{L},$
$U_{R},$ $D_{R},$ $L,$ $E_{R}$ \ correspond
to the chiral superfields, which
contain ordinary quarks and
leptons. \, $H_{1}$ \, and \, $H_{2}$ \, are MSSM Higgs fields.

\begin{center}
\begin{tabular}{|c|c|c|c|c|c|c|c|}
\hline
Superfield & $Q_{L}$ & $U_{R}$ & $D_{R}$ & $L$ & $E_{R}$ & $H_{1}$ & $H_{2}$
\\ \hline
$Q$ & $-\frac{3}{5}n$ & $-\frac{3}{5}n$ & $\frac{4}{5}n$ & $\frac{4}{5}n$ & $%
-\frac{3}{5}n$ & $-\frac{1}{5}n$ & $\frac{6}{5}n$ \\ \hline
\end{tabular}
\end{center}

\vspace{0.2cm} {\small {{\bf Table~1} The \ $U(1)_{A}$
charges corresponding
to the quarks, leptons and Higgs superfields defined up to the
multiplication by arbitrary number $n$.}}

\vspace{0.2cm}

For our analyses the ratio between
anomalous charges and their exact values
are not important. To simplify our analyses
let us denote corresponding
charges of ~MSSM Higgs \ $(H_{1}$ \
and \ $H_{2})$ \ as \ $q$ \ and \ $p$
\, respectively.

According to the previous
analyses (possible suppression of $~\mu $--term)
the scalar potential for the neutral
components of ~MSSM Higgs field
contains just the contributions
from $D$--terms corresponding to anomalous
and non--anomalous gauge symmetry and
is the following:
\begin{equation}
V=\frac{g_{A}^{2}}{4}(\xi +q|H_{1}^{0}|^{2}+p|H_{2}^{0}|^{2})^{2}+\frac{%
g_{W}^{2}+g_{Y}^{2}}{4}(|H_{1}^{0}|^{2}-|H_{2}^{0}|^{2})^{2},  \label{pot}
\end{equation}
where \ $g_{W}$, \, $g_{Y}$ \, and $g_{A}$ \, correspond
to gauge coupling
constant of \ $SU(2)_{W}\otimes U(1)_{Y}\otimes U(1)_{A}$\ gauge symmetry.
The absolute minimum of
the potential (\ref{pot}) is achieved for the
following vacuum expectation values of MSSM Higgs fields:
\begin{equation}
<|H_{1}^{0}|^{2}>=<|H_{2}^{0}|^{2}>=-\frac{\xi }{q+p}.  \label{vev}
\end{equation}
Based on the discussion given in the
previous subsection the value of \
$\sqrt{\xi }$--terms in the low--scale string
theory can be of the order of \, 
$~10^{2}$ GeV. So, under the natural assumption that $p + q$ is of order of the unity we can conclude
from eq. (\ref{vev}) \ that due to the
existence of anomalous \ $U(1)_{A}$ \ symmetry it
is possible to break SM
gauge symmetry and get the correct values of
masses of gauge bosons, Higgs
fields, quarks and leptons (as well as of their superpartners) without
having the $\mu $--term in the theory and
without extension of the particle
spectrum. However, in the absence of the
$\mu $--term it is easy to see that
 one of the charginos and one of the
neutralinos from gaugino and higgsino sector become massless.

The simplest way to avoid this
problem is to introduce additional superfield
\ $P$, \ which is a singlet under
the $SU(2)_{W}\otimes U(1)_{Y}$ symmetry and
has \ $h=-(p+q)$ \ charge under the
anomalous \ $U(1)_{A}$ \ symmetry. This
allows the existence of the following expression in the superpotential:
\[
W=\frac{k}{\sqrt{2}}H_{1}H_{2}P.
\]
The corresponding potential for \ $P$ ~field and
neutral components of ~MSSM
Higgs fields is:
\begin{eqnarray}
V &=&\frac{g_{A}^{2}}{4}(\xi
+q|H_{1}^{0}|^{2}+p|H_{2}^{0}|^{2}+h|P|^{2})^{2}+
\frac{g_{W}^{2}+g_{Y}^{2}}{4}
(|H_{1}^{0}|^{2}-|H_{2}^{0}|^{2})^{2}  \nonumber  \label{pot1} \\
&&+\frac{k^{2}}{2}
(|H_{1}^{0}H_{2}^{0}|^{2}+|H_{1}^{0}P|^{2}+|H_{2}^{0}P|^{2}).
\end{eqnarray}
From the extremum condition we have the
following system of linear equations:
\begin{eqnarray}
(g\,q^{2}+g^{\prime })|H_{1}^{0}|^{2}+(q\,p\,q-g^{\prime
}+k^{2})|H^{0}|_{2}^{2}+(g\,h\,q+\lambda )
|P|^{2}+g\,\xi \,q &=&0,  \nonumber
\label{extr} \\
(g\,p\,q-g^{\prime }+k^{2})|H_{1}^{0}|^{2}+
(g\,p^{2}+g^{\prime})
|H_{2}^{0}|^{2}+(g\,h\,p+k^{2})
|P|^{2}+g\,\xi \,p &=&0,  \nonumber \\
(g\,q\,h+k^{2})|H_{1}^{0}|^{2}+
(g\,p\,h+k^{2})|H_{2}^{0}|^{2}+g\,h^{2}
\,|P|^{2}+g\,\xi \,h &=&0,
\end{eqnarray}
where \ $g\equiv g_{A}^{2}$ \
and \ $g^{\prime }\equiv g_{W}^{2}+g_{Y}^{2}$.
It is clear that the solution
which corresponds to non-zero VEV of \
$|H_{1}^{0}|^{2},$ \ $|H_{2}^{0}|^{2}$ \ and \ $|P|^{2}$ \ fields is
proportional to \ $\xi $--term
and, as we noted before, \ $\sqrt{\xi }\sim O$
($10^{2}$GeV).\ Hence, in this
way it is possible to generate correct VEV
for SM Higgs fields. The non--zero
VEV of \, $P$ \, field
generates effectively the \ $\mu $--term in
the superpotential. This is
equivalent to the arising of \ $\mu $--term in
gaugino--higgsino sector and in
this way all neutralinos and charginos become massive.

Let us note, that in this scenario supersymmetry is spontaneously
broken as well
and the masses of superpartners  of the Standard Model particles
are of order of $\xi$.
Thus, in this case we have the
self--consistent picture for supersymmetric SM.

\section{The model 2}

\subsection{The possible generation of $~\mu $--term}

The second model that we are consider suggests
the \, $\mu $--term to be of
some dynamical origin, namely it is a
vacuum expectation value of some extra
fields living in the "hidden sector".
As we will show this is the
necessary condition for the realization of
mechanism of electroweak symmetry
breaking which is based on the
existence of anomalous \ $U(1)_{A}$ \ symmetry.

Let us explain the possibility of the
appearance of the \ $\mu $--term in
the four dimensional effective action.
One way to observe this is to
consider the \ $d=4$\ $N=1$ \ supergravity
lagrangian, which can arise after
the appropriate compactification of
type {\bf IIB} or heterotic strings. The
\ $d=4$ \, $N=1$ \ supergravity
lagrangian is a functional of three
independent functions \ $K,$ $f,$ $f^{\prime }$ \ of
the dilaton, modulus,
visible and hidden sector chiral
superfields as well as of the vector
superfields which correspond to
the gauge interactions. The functions \ $K$
~and ~ $f$ ~contribute to the
K\"{a}hler potential, while $f^{\prime }$ \, is
the gauge kinetic function. The
problem of finding the exact expression of
these functions in the superstring
theory is still open although various
string vacua give a certain
restrictions on them \cite{ibq}, \cite{KK}.

For the simplification of our analyses let us discard
 the dependence of
K\"{a}hler potential on the modulus fields.
The part of the \ $d=4$
$N=1$\ supergarvity K\"ahler potential describing
the interaction of gravity with
chiral matter superfields \ $y^{a}$ \ and
the chiral superfields \ $z^{i}$,
\,  coming from the hidden sector, is expressed as:
\begin{equation}
G=K(\Phi ,\Phi ^{+})+\ln |f(\Phi )|^{2},  \label{Lagr}
\end{equation}
where \ $K$\ depends
on \ $\Phi \equiv (y^{a},$ $z^{i})$\, and its conjugate
$\Phi ^{+}$ and \ $f$\ ~is a
holomorphic function of \ $\Phi $ .

As it was shown in \cite{GM} one can
take the functions \ $K(\pi ,\pi^{+},y,y^{+})$ \
and \ $f(\pi ,y)$ \ ($\pi ^{i}\equiv z^{i}/M_{Pl}$) to be
of the following form:
\begin{equation}
f(\pi ,y)=M_{Pl}^{2}f^{(2)}(\pi )+
M_{Pl}f^{(1)}(\pi )+f^{(0)}(\pi ,y),
\end{equation}
\begin{equation}
K(\pi ,\pi ^{+},y,y^{+})=M_{Pl}^{2}
d^{(2)}(\pi ,\pi ^{+})+M_{Pl}d^{(1)}(\pi
,\pi ^{+})+d^{(0)}(\pi ,\pi ^{+},y,y^{+}),
\end{equation}
where
\begin{equation}
d^{(0)}=y^{a}\delta _{a}^{b}y_{b}^{+}+(\sum_{i}c_{m}^{\prime }
(\pi ,\pi^{+})
g_{m}^{(2)}(y)+h.c.),  \label{Kahler}
\end{equation}
\begin{equation}
f^{(0)}=\sum_{i}c_{n}(\pi )g^{(3)}(y).
\end{equation}
 The functions \ $g^{(3)}(y)$ \
and \ $g_{m}^{(2)}(y)$ \ are trilinear
and bilinear polynoms in $y$ fields. The  $\mu $--terms
for the matter superfields  are generated in d=4
N=1 potential
\begin{equation}
V=e^{G/M_{Pl.}^{2}}\left( \frac{\partial G}{\partial \Phi ^{A}}
\left( {\frac{\partial^{2}G}{\partial \Phi^{A}\partial
\Phi _{B}^{+}}}\right) ^{-1}\frac{%
\partial G}{\partial \Phi_{B}^{+}}-3\right) \quad +D-terms,
\end{equation}
via the condensate of "hidden sector" fields, namely
\begin{equation}
\mu _{m}=m\langle (1-\rho_{i}\frac{\partial }{\partial \pi ^{+}_i}
c_{m}^{\prime }(\pi ,\pi^{+})\rangle ,  \label{mass}
\end{equation}
with
\begin{equation}
\rho _{i}=\frac{\partial ^{2}d^{(2)}(\pi ,\pi ^{+})}{\partial \pi
^{i}\partial \pi _{j}^{+}}\frac{\partial }{\partial \pi ^{j}}(\ln
f^{(2)}(\pi )+d^{(2)}(\pi ,\pi ^{+})),
\end{equation}
\begin{equation}
m=\langle \exp (d^{(2)}(\pi ,\pi ^{+})/2)f^{(2)}(\pi )\rangle
\end{equation}
the later being the gravitino mass.

At present we are interested in how the
mass terms for the Higgs fields can
appear in the effective 4--dimensional theory.
Hence, we concentrate on the
part of (\ref{Kahler}) which contains
$H_{1}$ and $H_{2}$ superfields.
Obviously the $\mu $--term is still non--invariant
with respect to the
anomalous \ $U(1)_{A}$ \ gauge symmetry (for the Higgs fields we are
considering  \ $p\neq -q$ ). The situation
can be improved if we assume
the presence of the fields $\lambda ^{i}$ charged
under the anomalous \
$U(1)_{A}$ \ in the "hidden sector" $\pi $ and
contributing into the
lagrangian via the last term of equation (\ref{Kahler}):
\begin{equation} \label{Gamma}
\Gamma (\lambda _{i},\lambda _{i}^{+},y,y^{+})=\sum_{i}{(\lambda
_{i}^+)}^{N}H_{1}H_{2}+h.c.\ .
\end{equation}
This leads to the desired result if t
he fields $\lambda _{i}$ have charges
equal to \ $-(p + q)/N$ \ under \ $U(1)_{A}$ .
In this case the parameter \, $\mu$ \, will have the same
form given by (\ref{mass}) but with the holomorphic
function \ $c^{\prime }(\lambda^+)=\sum_{i}
{(\lambda_{i}^+)}^{N} $ \ in the right-hand side.

\subsection{The model}

Let us analyze the four dimensional Higgs
potential within the given
scenario. Using the invariance of $N=1$ supergravity
lagrangian under the
K\"{a}hler transformations \cite{GM} one can
formulate the theory in terms of a single
function  --~ the
superpotential, which contains the
term:
\begin{equation}
W^{\prime }=\frac{\mu }{\sqrt{2}}\,H_{1}\,H_{2}. \label{WWW}
\end{equation}
 Along with the anomalous and non--anomalous
D--terms the following
potential for the neutral components of the supersymmetric
SM Higgs fields is generated:
\begin{equation}
V=\frac{g}{4}(\xi +q|H_{1}^{0}|^{2}+
p|H_{2}^{0}|^{2})^{2}+\frac{g^{\prime }}
{4}(|H_{1}^{0}|^{2}-|H_{2}^{0}|^{2})^{2}+\frac{\mu ^{2}_1}{2}
\,(|H_{1}^{0}|^{2}+|H_{2}^{0}|^{2}),
\end{equation}
with \ $g\equiv g_{A}^{2}$,
$\mu_1^2 = \mu^2 + m^2$ \ and \
$g^{\prime }\equiv g_{W}^{2}+g_{Y}^{2}$.
Note that the presence of the term
(\ref{Gamma}) in the K\"ahler potential
 does not change the form (\ref{FI}) of the
considered D--terms
according to (\ref{fi}).
Let us note, that the superpotential (\ref{WWW})
generates also the term $B m \mu H_1 H_2$ with
$B=\frac{(2 - \rho \partial / \partial \pi^+) c^\prime}
{(1 - \rho \partial / \partial \pi^+) c^{\prime}}$
in d=4 N=1 SUGRA lagrangian, 
which however can be neglected and we do not consider it for
simplicity.

Solutions that correspond to the
non--zero VEV of SM Higgs fields are the
following:
\begin{equation}
<|H_{1}^{0}|^{2}>=-\frac{\mu ^{2}_1(2\,g^{\prime }+g\,p\,(p-q))+\xi
\,g\,g^{\prime }\,(p+q)}{g\,g^{\prime }\,(p+q)^{2}}  \label{sol1}
\end{equation}
and
\begin{equation}
<|H_{2}^{0}|^{2}>=-\frac{\mu ^{2}_1(2\,g^{\prime }+g\,q\,(q-p))+\xi
\,g\,g^{\prime }\,(p+q)}{g\,g^{\prime }\,(p+q)^{2}}.  \label{sol2}
\end{equation}
In order to have the physically
accepted result the right-hand sides of (\ref
{sol1}) and (\ref{sol2}) must be
positive. It is easy to see that it can be
realized with the  help of \ $\xi $--term if 
we suppose that the absolute value of
first term in eq. (\ref{sol1}) and (\ref{sol2}) is less then second one.
Thus, the anomalous $U(1)_{A}$ symmetry
again plays a crucial role for the
formation of correct values of \, VEV \, for \, SM
\, Higgs fields.

It is easy to see that VEV of \ $D_{A}$ - terms,
that corresponds to
anomalous \ $U(1)_{A}$ \ gauge symmetry in our
scenario, is different from
zero, that indicates the supersymmetry
breaking. In this case extra contribution to
the scalar masses arises for fields
having nontrivial charges under
considered $U(1)_{A}$ symmetry. From the
Eqs. (\ref{FI}), (\ref{sol1}) and (\ref{sol2}) one obtains:
\begin{equation} 
\Delta m_{i}^{2}=Q_{i}\frac{4\,\mu ^{2}_1}{p+q}.  \label{dtcon}
\end{equation}
This contribution can be useful for
the solution of supersymmetric flavor
problem, namely the relevant
phenomenology of this kind was discussed in
\cite{PD,tk}. The point is that
the value of $F$ -- terms which are nonzero
as well and give the contribution
to the squark and slepton masses 
are suppressed
by the factor $\frac{\xi}{M_{Pl}^2}$
and therefore are small with respect to the contribution (\ref{dtcon}). 
Notice that the same kind of contribution
to the scalar masses
takes place in the model considered in the previous
Section as well.

Hence, we have proven that  embedding
of the MSSM into the \ $d=4$
\, $N=1$ \ superstring models
when string scale is of the order of \
TeV the electroweak symmetry breaking mediated by the
Fayet--Illiopoulos
term corresponding to an
anomalous \ $U(1)_{A}$ \ $D_{A}$--term can take place.

\section{Summary}

In this letter we have studied a possible
influence of anomalous \ $U(1)_{A}$
\ symmetry on the violation of SM gauge
symmetry in the supersymmetric case.
Our results show that the value of
Fayet--Illiopoulos \ $\xi $--term is
crucial for the generation of
correct VEV of Higgs fields when having the
low--scale superstring theory. The only
important requirement on the value
of parameter $\mu $ to have the self--consistent
picture of electroweak
symmetry breaking is $|\mu |<|\xi |$.
In contrast with the other models
discussed in the literature the order of
parameter $\mu $ is unimportant for
the formation of correct vacuum VEVs of the Higgs
fields and can be fixed
from the low bound of chargino--neutralino section.
In both suggested models
we have found the extra contribution
from VEV of \, $D_{A}$--term. This result
cane be used for the solution of the supersymmetric 
flavor problem.

The models considered in
this paper have left several unresolved questions
both from the theoretical
and phenomenological points of view. Namely, it
would be interesting to obtain these
models using the string theory
computation and  to
examine whether \ $SU(2)_{W} \otimes U(1)_{Y}$ \
symmetry breaking can take
place in the known four dimensional  superstring
models. The latter procedure
can  require the inclusion of
several    $\xi^i$   terms corresponding to
 anomalous 
\ $U(1)^i_A$ \ symmetries present in type {\bf IIB}
superstring theory.

\noindent {\bf Acknowledgments} We are grateful to Zurab Berezhiani, 
Durmush Demir, Gia
Dvali, Edi Gava, Alexei Gladyshev,
Dimitri Gorbunov, Elias Kiritsis, and
Goran Senjanovic for interest to this work and
helpful discussions. Work of
M.T. was supported in part by the Russian Foundation of Fundamental
Research, under the grant 99-02-18417.

\end{document}